\documentclass[aps,prd,preprint,a4paper,showpacs,showkeys,superscriptaddress]{revtex4-1}

\usepackage{latexsym}
\usepackage{amsmath,amssymb}
\usepackage{graphicx}
\usepackage{xcolor}

\usepackage{acronym}

\usepackage{hyperref}     
\hypersetup{colorlinks,%
  citecolor=blue,%
  linkcolor=cyan,%
  pdftex}

\acrodef{AMPS}[AMPS]{Almheiri, Marolf, Polchinski, and Sully}
\acrodef{CGHS}[CGHS]{Callan-Giddings-Harvey-Strominger}
\acrodef{RST}[RST]{Russo-Susskind-Thorlacius}
\acrodef{GEMS}[GEMS]{global embedding in Minkowski spacetime}
\acrodef{AdS}[AdS]{anti-de Sitter}
\acrodef{SAdS}[SAdS]{Schwarzschild anti-de Sitter}

\begin{document}


\title{Proper temperature of the Schwarzschild AdS black hole revisited}

\author{Myungseok Eune}%
\email[]{eunems@smu.ac.kr}%
\affiliation{Department of Civil Engineering, Sangmyung
  University, Cheonan, 31066, Republic of Korea}%

\author{Wontae Kim}%
\email[]{wtkim@sogang.ac.kr}%
\affiliation{Department of Physics, Sogang University, Seoul, 04107,
  Republic of Korea}%

\date{\today}

\begin{abstract}
  The Unruh temperature calculated by using the global embedding of
  the Schwarzschild AdS spacetime into the Minkowski spacetime was
  identified with the local proper temperature; however, it became
  imaginary in a certain region outside the event horizon. So, the
  temperature was assumed to be zero of non-thermal radiation for that
  region.  In this work, we revisit this issue in an exactly soluble
  two-dimensional Schwarzschild AdS black hole and present an
  alternative resolution to this problem in terms of the Tolman's
  procedure.  However, the process appears to be non-trivial in the
  sense that the original procedure assuming the traceless
  energy-momentum tensor should be extended in such a way that it
  should cover the non-vanishing case of the energy-momentum tensor in
  the presence of the trace anomaly.  Consequently, we show that the
  proper temperature turns out to be real everywhere outside the event
  horizon without any imaginary value, in particular, it vanishes at
  both the horizon and the asymptotic infinity.
\end{abstract}



\maketitle


\section{Introduction}
\label{sec:intro}

One of the most outstanding theoretical results in quantum mechanics
of black holes would be Hawking radiation~\cite{Hawking:1974rv,
  Hawking:1974sw}. On general grounds, the thermal distribution of
Hawking radiation could be characterized by the two black hole
temperatures associated with the observers.  One is the fiducial
temperature measured by a fixed observer who undergoes acceleration,
which is usually given as the redshifted Hawking temperature at a
finite distance outside the event
horizon~\cite{Weinberg:1972book}. The other is the proper temperature
measured by a freely falling observer from rest, which is expressed by
the Tolman temperature~\cite{Tolman:1930zza, Tolman:1930ona}.

Surprisingly, the fiducial temperature takes the same form as the
proper temperature even though the respective observers belong to
different frames. At first glance, one might conclude that the
equivalence principle could be violated, particularly at the horizon
from the fact that the Tolman temperature could be divergent there.
However, it is worth noting that the equivalence principle could be
restored just at the horizon as seen from the calculations using the
particle detector method~\cite{Singleton:2011vh}.  This fact could
also be confirmed by showing that the Tolman temperature could vanish
effectively at the horizon~\cite{Gim:2015era}.

As a matter of fact, the Tolman temperature for the freely falling
observer should be modified effectively, so that its behavior could be
shown to be definitely different from that of the fiducial
temperature~\cite{Gim:2015era}.  Recently, a similar argument for the
proper temperature~\cite{Barbado:2016nfy} could also be obtained
from a different point of view by clarifying the Hawking
effect~\cite{Hawking:1974rv} and the Unruh effect~\cite{Unruh:1976db}.
Note that all these arguments are for asymptotically flat black holes,
and it appears to be natural to ask how to get the proper
temperatures in asymptotically non-flat spacetimes such as the
\ac{SAdS} black hole.

Regarding the calculations of the proper temperature in the \ac{SAdS}
black hole, there has been pioneering works employing the \ac{GEMS}
approach, where an accelerating observer in a higher-dimensional
Minkowski spacetime perceives thermal radiation characterized by the
Unruh temperature which will be identified with the proper temperature
in the original spacetime~\cite{Deser:1997ri, Deser:1998xb}; however,
the proper temperature suffers from an imaginary value. So, it was
claimed that the imaginary valued proper temperature would indicate
non-thermal radiation~\cite{Brynjolfsson:2008uc}.  The non-thermal
condition to evade the imaginary temperature seems to be somehow
\textit{ad hoc}.  Obviously, the temperature could be made real in the
near horizon limit when a reduced embedding was
used~\cite{Banerjee:2010ma}.

Now it raises a question: is there any other way to resolve this
imaginary value problem for the proper temperature in the \ac{SAdS}
black hole?  In fact, there is another way to calculate the proper
temperature directly, which is the old-fashioned but clear Tolman
procedure~\cite{Tolman:1930zza, Tolman:1930ona}, which might provide a
plausible solution to this question.  However, this approach appears
to be conceptually non-trivial in the sense that the conventional
Tolman temperature derived from the conventional Stefan-Boltzmann law
rests upon the traceless condition of the energy-momentum tensor.  If
one were to study the proper temperature on the background of the
asymptotically \ac{AdS} spacetimes, the traceless condition for the
energy-momentum tensor should be released in order to take into
account the non-vanishing trace of the quantized energy-momentum
tensor in the presence of the trace anomaly.

In this work, we would like to revisit the proper temperature of the
\ac{SAdS} black hole and show how to get the well-defined real-valued
proper temperature.  In essence, we shall obtain a modified
Stefan-Boltzmann law, which is commensurate with the presence of the
non-vanishing trace of the energy-momentum tensor. Then, from the
modified Stefan-Boltzmann law, we shall derive an effective Tolman
temperature and obtain the desired result.  In fact, such a
modification of the Stefan-Boltzmann law has already been applied to
various models: thermodynamics of particle physics in flat
spacetime~\cite{Boyd:1996bx}, thermodynamics of black hole in curved
spacetime~\cite{Gim:2015era}, and warm inflation models in cosmology
\cite{Gim:2016uvv}, so that some puzzling problems have been
successfully resolved.

Our calculations will be done in a two-dimensional amenable model in
order to solve exactly without losing any essential physics.  In
Sec.~\ref{sec:gems}, the proper temperature will be obtained in the
two-dimensional \ac{SAdS} black hole by using the
\ac{GEMS}~\cite{Deser:1997ri, Deser:1998xb, Brynjolfsson:2008uc} in
the self-contained manner in comparison with our result.  As expected,
we find that the imaginary temperature is unavoidable in a certain
region.  Next, in Sec.~\ref{sec:SB}, we will calculate the proper
temperature from the Tolman's procedure~\cite{Tolman:1930zza,
  Tolman:1930ona} by releasing the traceless condition for the
energy-momentum tensor.  We shall show that the proper temperature
for the \ac{SAdS} black hole turns out to be real everywhere outside
the horizon without any imaginary value, so that it becomes smooth
without any cusp.  Summary and discussion will be given in
Sec.~\ref{sec:discussion}.

\section{Proper temperature from the GEMS}
\label{sec:gems}

We recapitulate how the proper temperature for the two-dimensional
\ac{SAdS} black hole could be derived from the framework of the
\ac{GEMS} employed in Ref.~\cite{Brynjolfsson:2008uc}.  Let us start
with the two-dimensional \ac{SAdS} black hole described by
\begin{align}
  ds^2 = -f(r) dt^2 + \frac{dr^2}{f(r)},  \label{line.element}
\end{align}
where $f(r) = 1 - 2M/r + r^2/\ell^2$. The metric element can be
rewritten as
\begin{align}
  \label{SAdS.h}
  f(r) = \frac{1}{\ell^2 } \left(1-\frac{r_h}{r}\right) (r^2 + r r_h + r_h^2 + \ell^2),
\end{align}
where $r_h$ is the horizon of the black hole and the mass is related
to the horizon as
\begin{align}
  \label{SAdS:mass}
  M = \frac{r_h}{2\ell^2} (r_h^2 + \ell^2).
\end{align}
And the surface gravity is also given by
\begin{align}
  \kappa &=  \frac{f'(r_h)}{2} = \frac{3r_h^2 + \ell^2}{2\ell^2
           r_h}, \label{kappa:SAdS}
\end{align}
where the prime denotes the derivative with respect to $r$.

Performing the global embedding of the \ac{SAdS} spacetime into the
higher dimensional Minkowski spacetime, a free-fall observer on the
\ac{SAdS} black hole could be identified with the accelerated observer
in the higher dimensional Rindler spacetime~\cite{Deser:1997ri,
  Brynjolfsson:2008uc}, so that the Rindler observer could find the
Unruh temperature as~\cite{Unruh:1976db}
\begin{align}
  T &= \frac{a}{2\pi}, \label{T:unruh}
\end{align}
where $a$ is the proper acceleration of the observer in the higher
dimensional Minkowski spacetime.

The higher dimensional Minkowski spacetime can be obtained by the
following transformation~\cite{Deser:1998xb}
\begin{align}
  X^0 &= \kappa^{-1} \sqrt{f} \sinh \kappa t, \qquad X^1= \kappa^{-1}
        \sqrt{f} \cosh \kappa t, \qquad X^2 = r, \nonumber    \\
  X^3 &= \int dr \frac{\ell (r_h^2 + \ell^2 )}{3r_h^2 + \ell^2}
        \sqrt{\frac{r_h(r^2 + r r_h + r_h^2)}{r^3(r^2 + r r_h + r_h^2
        + \ell^2)}}, \nonumber \\
  X^4 &= \int dr \frac{1}{3r_h^2 + \ell^2}
        \sqrt{\frac{(9r_h^4 + 10 r_h^2 \ell^2 + \ell^4)(r^2 + r r_h +
        r_h^2)}{r^2 + r r_h + r_h^2 + \ell^2}}, \label{X4:SAdS:gems}
\end{align}
where the line element is $ds^2 = \eta_{IJ} dX^I dX^J$ with
$\eta_{IJ} = {\rm diag}(-1,1,1,1,-1)$. In this spacetime, the square
of the proper acceleration is calculated as
\begin{align}
  a^2 &= \eta_{IJ} a^I a^J  \notag \\
      &= \frac{[2 + (3+c^2)x + (1+c^2) x^3][-2 + (1+c^2)
        (1+x)x]}{4\ell^2 [1 + x + (1+c^2)x^2]},  \label{SAdS:gems:accel}
\end{align}
where $x = r_h/r$ and $c=\ell/r_h$.  Substituting this
acceleration~\eqref{SAdS:gems:accel} into Eq.~\eqref{T:unruh}, the
Unruh temperature regarded as the proper temperature is obtained as
\begin{align}
  T &= \frac{1}{4\pi\ell} \frac{\sqrt{[2 + (3+c^2)x + (1+c^2) x^3][-2 + (1+c^2)
      (1+x)x]}}{\sqrt{1 + x + (1+c^2)x^2}}.  \label{SAdS:gems:T}
\end{align}
The squared proper temperature is positive $r<r_c$, while it is
negative for $r > r_c$ where the critical radius is given by
$r_c = (r_h^2 + \ell^2 + \sqrt{9r_h^4 + 10 r_h^2 \ell^2 + \ell^4})
/(4r_h)$.  In particular, the squared temperature at the horizon
becomes
\begin{align}
  T^2(r_h) = \frac{1}{4\pi^2} \left[ \frac{\ell ^2}{2 r_h^4}+\frac{3}{3 r_h^2+\ell ^2} \right],
  \label{SAdS:gems:accel:horizon}
\end{align}
which is positive finite.  By the way, at the asymptotic infinity, the
squared temperature takes the form of
\begin{align}
  T^2(\infty) \to -  \frac{1}{4\pi^2} \left[ \frac{1}{\ell^2} +
  \frac{\left(r_h^2+\ell ^2\right)^2}{2 r_h^4 \ell ^2}
  \right], \label{SAdS:gems:accel:infinity}
\end{align}
which is negative. Thus one can find that the proper temperature
becomes imaginary for $r > r_c$.  In fact, it was claimed that the
imaginary proper temperature would indicate non-thermal
radiation~\cite{Brynjolfsson:2008uc}, and the proper temperature
was assumed to be zero for $r > r_c$. It means that there would appear
a cusp at $r_c$ for the temperature curve.  In the next section, we
will find another way to resolve this imaginary value problem by
directly calculating the proper temperature through the Tolman's
procedure.

\section{Proper temperature from the Tolman procedure}
\label{sec:SB}

We calculate the temperature measured by a freely falling observer
released from rest on the \ac{SAdS} black hole.  Here, we shall
release the traceless condition employed in the conventional
formulation of the Tolman temperature~\cite{Tolman:1930zza,
  Tolman:1930ona} in order to get the effective Tolman temperature.

Let us start with the proper velocity of a particle obeying the
geodesic equation of motion
\begin{align}
  u^\mu = \left(\frac{\alpha}{f}, - \sqrt{\alpha^2 - f}
  \right),  \label{proper.velocity}
\end{align}
where $\alpha$ is an integration constant. The freely falling observer is
released at $r=r_0$ with the zero velocity, and then the integration
constant can be determined by $\alpha = \sqrt{f(r_0)}$.  In the
conformal gauge of $ds^2 = - e^{2\sigma} dx^+ dx^-$ with
$e^\sigma = \sqrt{f(r)}$, the proper velocity~\eqref{proper.velocity} is
rewritten as
\begin{align}
  u^\pm &= \frac{\sqrt{f(r_0)} \mp \sqrt{f(r_0) - f(r)}}{f(r)}, \label{u:pm}
\end{align}
and the unit normal vector is chosen as $n^+ = u^+$ and $n^- = -u^-$,
where they satisfy $u^\mu u_\mu = -1$, $u^\mu n_\mu = 0$, and
$n^\mu n_\mu =1$. We consider a free-fall frame at $r=r_0$, and then
Eq.~\eqref{u:pm} reduces to
\begin{align}
  u^\pm &= \frac{1}{\sqrt{f(r_0)}}, \label{u:pm:FFAR}
\end{align}
where $r_0$ will be replaced by $r$ for a simple notation hereafter.

On the other hand, it has been well-known that Hawking radiation is
related to the trace anomaly~\cite{Christensen:1977jc}, which means
that the traceless condition of the energy-momentum tensor should be
released in the thermodynamic black hole system.  Explicitly, the
trace anomaly for a single scalar field in two dimensions is given by
\begin{align}
   T^\mu_\mu  = \frac{1}{24\pi} R,  \label{trace.anomaly}
\end{align}
where the scalar curvature is written as $R = -f''$ for the line
element~\eqref{line.element}. From the trace
anomaly~\eqref{trace.anomaly} with the help of the conservation law
for the energy-momentum tensor, the energy-momentum tensor is written
as~\cite{Callan:1992rs}
\begin{align}
  T_{\pm\pm}
  &= \frac{1}{96\pi} \left[ff'' - \frac12
    (f')^2 + t_\pm \right], \label{T:++}\\
  T_{+-}
  &= \frac{1}{96\pi} ff''. \label{T:+-}
\end{align}
where $t_\pm$ reflect the nonlocality of trace anomaly.  For the
Hartle-Hawking-Israel state~\cite{Hartle:1976tp,Israel:1976ur},
$t_\pm$ are explicitly determined by
\begin{align}
  t_\pm &= \frac{1}{2}f'(r_h)^2, \label{bc}
\end{align}
where $t_+=t_-$ in thermal equilibrium and so the net
flux automatically vanishes.

Next, one can write down the energy density and pressure for the
freely falling observer as follows
\begin{align}
  \rho &=  T_{\mu\nu} u^\mu u^\nu \notag \\
       &= \frac{1}{96\pi f} \left[ 4 f f'' -  (f')^2 + t_+ + t_-
         \right], \label{energy.density}
\end{align}
and
\begin{align}
  p &=  T_{\mu\nu}  n^\mu n^\nu \notag \\
    &= \frac{1}{96\pi f} \left[-  (f')^2 + t_+ + t_-
      \right], \label{pressure}
\end{align}
respectively.  Then the explicit form of the proper energy density and
pressure are
\begin{align}
  \rho &= \frac{1}{96\pi f} \left[ 8 \left(1 -\frac{2
         M}{r}+\frac{r^2}{\ell ^2}\right) \left(-\frac{2 M}{r^3} +
         \frac{1}{\ell ^2}\right) -\left(\frac{2 M}{r^2}+\frac{2
         r}{\ell^2}\right)^2 + \left(\frac{2 M}{r_h^2}+\frac{2
         r_h}{\ell^2}\right)^2 \right], \label{energy.density:explicit}
\end{align}
and
\begin{align}
  p &=\frac{1}{96\pi f} \left[-\left(\frac{2 M}{r^2}+\frac{2
      r}{\ell^2}\right)^2 + \left(\frac{2 M}{r_h^2}+\frac{2
      r_h}{\ell^2}\right)^2 \right]. \label{pressure:explicit}
\end{align}
Note that the proper energy density is negative finite at the horizon
such as $\rho(r_h) \to -(1/12\pi r_h^2)$ and it is positive finite at
the asymptotic infinity, $\rho(\infty) \to 1/(24\pi\ell^2)$. So there
appears a special point to divide the region into the negative energy
density and the positive energy density. This kind of feature appears
even in asymptotically flat black
holes~\cite{Gim:2015era,Ford:1993bw}.  The attendant problem is how to
relate the positive and negative energy density to the corresponding
temperatures consistently. For this purpose, we have to extend the
conventional Stefan-Boltzmann law which is only valid for the positive
energy density.

We are now in a position to explain how to get the proper
temperature by using the modified Stefan-Boltzmann law to relate the proper
energy density to the proper temperature.  Let us start with the first law of
thermodynamics written as~\cite{Tolman:1930ona,Tolman:1930zza}
\begin{align}
  dU = TdS - pdV, \label{1st.law:thermodynamics}
\end{align}
where $U$, $S$, $V$, $T$, and $p$ are the internal energy, entropy,
volume, temperature, and pressure of a system, respectively. At a
fixed temperature, Eq.~\eqref{1st.law:thermodynamics} can be rewritten
as
\begin{align}
  \left(\frac{\partial U}{\partial V} \right)_T = T
  \left(\frac{\partial S}{\partial V} \right)_T - p, \label{1st.law:rewriting}
\end{align}
where $\left(\partial U /\partial V \right)_T$ is just the energy
density $\rho$. Using the Maxwell relation
\begin{align}
  \left(\frac{\partial S}{\partial V} \right)_T = \left(
  \frac{\partial p}{\partial T} \right)_V,
\end{align}
one can see that Eq.~\eqref{1st.law:rewriting} becomes
\begin{align}
  \rho = T \left(\frac{\partial p}{\partial T} \right)_V - p. \label{1st.law:diff.eq}
\end{align}
In addition to this, we note that the trace of the energy-momentum
tensor for a perfect fluid is generically non-vanishing, which is
given as
\begin{align}
  T^\mu_\mu = -\rho + p. \label{trace:fluid}
\end{align}
From Eqs.~\eqref{1st.law:diff.eq} and ~\eqref{trace:fluid}, we can
eliminate the pressure term and obtain the first order differential
equation as
\begin{align}
  \label{key}
  T \left(\frac{\partial \rho}{\partial T}\right)_V -2\rho = T^\mu_\mu,
\end{align}
where we used the fact that the trace anomaly is independent of
temperature~\cite{BoschiFilho:1991xz}.  Then, the energy density and
pressure are easily solved as
\begin{align}
  \rho &= \gamma T^2 - \frac12 T^\mu_\mu, \label{energy:thermo} \\
  p &= \gamma T^2 + \frac12 T^\mu_\mu, \label{p:thermo}
\end{align}
respectively, where $\gamma$ is an integration constant determined as
$\gamma =\pi/6$ for a scalar field.  Note that the modified
Stefan-Boltzmann law \eqref{energy:thermo} and \eqref{p:thermo}
naturally reduce to the conventional Stefan-Boltzmann
law~\cite{Tolman:1930zza,Tolman:1930ona}.
\begin{figure}[htb]
  \centering
  \includegraphics[width=0.7\linewidth]{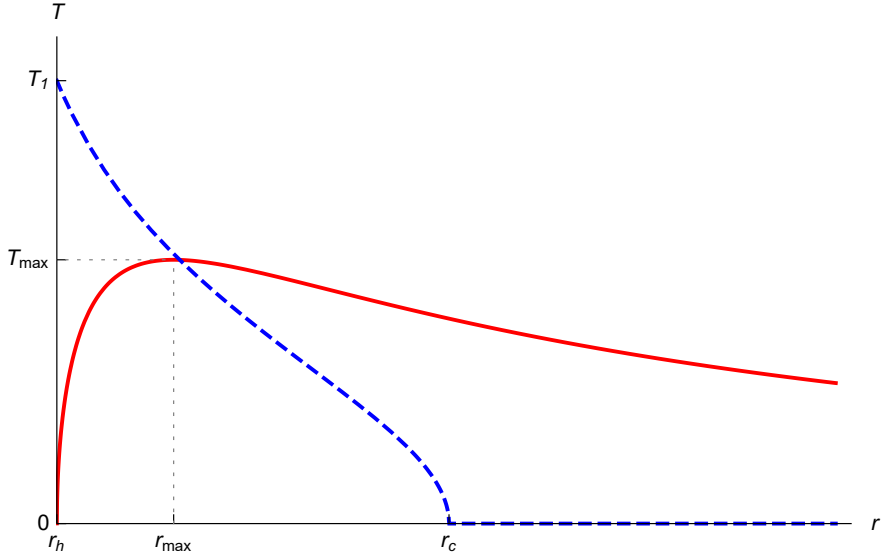}
  \caption{The two proper temperatures \eqref{SAdS:gems:T} and
    \eqref{T:square:SAdS:SB} are plotted by setting $r_h = 1$,
    $\ell = 2$, and $M=5/8$, for convenience. The dashed curve
    describes the behavior of the proper temperature from the
    \ac{GEMS}.  It has a cusp at $r_c \approx 3.266$, so that $T = 0$
    is imposed at $r > r_c$, where the imaginary temperature appears,
    and $T_1 = (2\pi r_h)^{-1}= (2\pi)^{-1}$ at $r =r_h$. The solid
    curve is the proper temperature based on the effective Tolman
    temperature, which is vanishing at both the event horizon $r_h$
    and the asymptotic infinity. It is always real and smooth, and
    reaches a peak of $T_{\rm max} \approx 0.0948$ at
    $r_{\rm max} \approx 1.675$.}
     \label{fig:temperature}
\end{figure}

Combining Eqs.~\eqref{trace.anomaly}, \eqref{energy.density}, and
\eqref{energy:thermo}, we obtain the squared effective Tolman
temperature as
\begin{align}
  T^2 &= \frac{1}{96\pi \gamma f} \left[2f f'' -  (f')^2 + t_+
        + t_- \right]. \label{square:T}
\end{align}
Plugging the boundary condition \eqref{bc} into Eq.~\eqref{square:T}, we
get the proper temperature as
\begin{align}
  T &= \frac{1}{4\pi \ell} \frac{x \sqrt{(1+c^2) (1-x) [3(3
      + 2x + x^2) + c^2 (1 + 2x + 3x^2)]}}{\sqrt{1 + x +
      (1 + c^2) x^2}}, \label{T:square:SAdS:SB}
\end{align}
where $x = r_h/r$ and $c =\ell/r_h$. The behavior of the
temperature~\eqref{T:square:SAdS:SB} is shown in
Fig.~\ref{fig:temperature}, where the proper temperature described by
the solid curve is real everywhere. In particular, it vanishes at both
the horizon and the asymptotic infinity, while it approaches the
maximum value at the critical radius of $r_{\rm max}$. The smooth
behavior of the effective Tolman temperature is in contrast to the
behavior of the proper temperature calculated by using the GEMS
method~\cite{Brynjolfsson:2008uc}, where one should require a
non-thermal condition by hand such as $T = 0$ for $r > r_c $ at which
the imaginary value appears.

On the other hand, if one takes the limit of $\ell \to \infty$ in
Eq.~\eqref{T:square:SAdS:SB} for the limit of the Schwarzschild black
hole, then one can find the proper temperature of
\begin{equation}
  T = \frac{1}{8\pi M}  \sqrt{(1-x)(1+2x+3x^2)}
\end{equation}
which exactly agrees with the previous result of the proper
temperature for the Schwarzschild black hole~\cite{Gim:2015era}.  So,
in this limit, the proper temperature vanishes at the horizon of
$x \to 1$ and it reproduces the Hawking temperature at the spatial
infinity of $x \to 0 $.

\section{Conclusion and discussion}
\label{sec:discussion}

In summary, the proper temperature on the background of the
two-dimensional \ac{SAdS} black hole has been investigated by using
the two different methods.  First, it was derived from the framework
of the \ac{GEMS}; however, it became imaginary for a certain region
such as $r > r_c$ as shown in Fig.~\ref{fig:temperature}.  So, it was
claimed that the imaginary temperature implies non-thermal radiation
in that region~\cite{Brynjolfsson:2008uc}.  In this work, we revisited
this issue by calculating the proper temperature straightforwardly
from the Stefan-Boltzmann law without resort to any indirect methods.
For this purpose, the conventional Stefan-Boltzmann law was extended
to the case of the non-vanishing trace of the energy-momentum tensor.
In essence, if one were to consider a black hole system with Hawking
radiation, then one should take into account non-trivial trace of the
energy-momentum tensor in the calculation of the Stefan-Boltzmann
law~\cite{Gim:2015era}.  Consequently, we could find the effective
Tolman temperature whose form is different from the conventional
Tolman temperature by the anomalous term in Eq.~\eqref{energy:thermo}.
The resulting effective Tolman temperature~\eqref{T:square:SAdS:SB} as
the proper temperature is always real and smooth without encountering
any imaginary value.

For the Schwarzschild black hole, it was shown that the equivalence
principle could be restored only at the
horizon~\cite{Singleton:2011vh}, and so it would be natural for the
proper temperature to vanish there~\cite{Gim:2015era,Barbado:2016nfy}.
The present calculation shows that the above feature could also be
found even in the \ac{SAdS} black hole.  This fact can be understood
by employing the Unruh effect.  For the large black hole, the
metric~\eqref{SAdS.h} is expressed by the Rindler metric in the near
horizon limit.  The Unruh effect tells us that the temperature is
given as $T_{\rm U}=a/2\pi$ near the horizon, where the acceleration
of the fiducial observer is
$a = M/( r^2 \sqrt{f})$~\cite{Unruh:1976db}.  It implies that the
free-fall observer could find the vanishing Unruh temperature, if the
frame is free from the acceleration.  In that sense, it seems to be
reasonable for the freely falling observer to find the vanishing
temperature at the horizon.  On the other hand, at the asymptotic
infinity, particles could not pass through the AdS boundary due to the
infinite potential.  In these regards the proper temperature vanishes
at the horizon and the asymptotic infinity.

Finally, one might wonder why the proper temperature calculated by
using the \ac{GEMS} method was different from the result derived from
the Tolman procedure. In the \ac{GEMS} method, the lower dimensional
black hole geometry is embedded into the higher dimensional Minkowski
spacetime classically, prior to the quantization of the
theory. However, the quantized theory formulated in the higher
dimensions might be inequivalent to the quantized theory in the lower
dimensions despite the classical equivalence.  It means that the
classical equivalence does not always warrant the quantum-mechanical
equivalence.  This speculation might deserve further attention.

\section*{Acknowledgments}
We would like to thank Y.\ Gim for exciting discussions.  W.\ Kim was
supported by the National Research Foundation of Korea (NRF) grant
funded by the Korea government (MSIP) (2014R1A2A1A11049571).


\bibliographystyle{JHEP}       

\bibliography{references}

\end{document}